# Generation of Motion of Drops with Interfacial Contact

Manoj K. Chaudhury[1]*, Aditi Chakrabarti[1] and Susan Daniel[2]

[1]Department of Chemical and Biomolecular Engineering, Lehigh University, Bethlehem, PA 18015
[2]School of Chemical and Biomolecular Engineering, Cornell University, Ithaca, NY 14850

**ABSTRACT**. A liquid drop moves on a solid surface if it is subjected to a gradient of wettability or temperature. However the pinning defects on the surface manifesting in terms of a wetting hysteresis, or a first order non-linear friction, limits the motion in the sense that a critical size has to be exceeded for a drop to move. The effect of hysteresis can, however, be mitigated by an external vibration that can be either structured or stochastic, thereby creating a directed motion of the drop. Many of the well-known features of rectification, amplification and switching that are generic to electronics can be engineered with such types of movements. A specific case of interest is the random coalescence of drops on a surface that gives rise to a self-generated noise. This noise overcomes the pinning potential diffusively, thereby generating a random motion of the coalesced drops. Randomly-moving coalesced drops themselves exhibit a purely diffusive flux when a boundary is present to eliminate them by absorption. With the presence of a bias, the coalesced drops execute a diffusive drift motion that can have useful application in various water and thermal management technologies.

* Correspondence should be addressed at mkc4@lehigh.edu

# 1. INTRODUCTION.

Transport of mass, momentum and heat constitute the basic paradigm of non-equilibrium systems. Fluid flows when there is a gradient of momentum density. Mass or heat flows when there is a gradient of chemical or thermal potential. There are also numerous examples of cross coupling of various forces[1] in nature that create immaculate specificity and control of transport. Biological systems[2] are also quite adept in rectifying noise to generate various types of directed transport processes. Scientists and engineers have paid a great deal of attention to the intricate artistry of nature in perfecting myriad of transport processes and succeeded in mimicking some of them in various technological settings over a period of several decades. Transport of liquid droplets on a surface is such an example, which provides a vast diversity of directed transports. The field has become particularly enriched by its immense importance in microfluidics[3], thermal[4] and water management[5,6] technologies. We learned over the years how the motion of drops can be induced when it is in contact with a surface by using surface energy gradient[7], a chemical interaction[8], and a thermal gradient[9].

One common feature of a drop on a surface is that its three phase contact line is pinned by the defects on the surface[10-12]. Although, such a pinning can be a detriment to droplet motion, we learned how to use them to our advantage in breaking the symmetry[13,14] of an externally imposed periodic force that is, conceptually, not too far from the way electrical engineers tame the oscillating electrons with a diode. Liquid drops also exhibit various spherical harmonic modes of vibration[15,16] leading to shape fluctuations that can amplify the droplet speed at specific resonance frequencies[17]. Furthermore, we also learned that an external vibration is not absolutely necessary to obtain directed motion. Drops growing on a surface due to the condensation of vapor can coalesce with each other thus leading to a shape fluctuation and an internally generated noise. The center of mass of the coalesced drops executes a self-avoiding random walk[18,19], which can be rectified by a chemical or a thermal bias to generate directed transport[4]. Even in the absence of a field, these randomly moving drops lose their identity when they come in contact with a continuous liquid film, thus leading to a motion[20] akin to the diffusive flux of a Brownian particle with an absorbing wall. The subject is vast and it is not possible to review all the advancements that took place in recent years in this short article. This article is meant to review mainly the research carried out in our group spanning a period of nearly two decades by highlighting the physics of symmetry breaking that govern various drop transport processes.

## 2. MOTION DUE TO CHEMICAL GRADIENT.

We begin this article by paying tribute to a recently deceased molecular biologist M. Steinberg, who proposed[21] half a century ago the hypothesis of differential adhesion as a means to explain the specificity of cell organization and tissue growth. This hypothesis showed a remarkable insight in which cellular motion was related to an adhesion gradient and in which a geometric mean combining rule was pursued for the adhesion of dissimilar materials on probabilistic grounds, this being independent of a similar rule[22] known to surface physicists based on molecular polarizations. Subsequently, Carter[23] explored the idea of differential adhesion in the context of the haptotactic migrations of cells, who noted that cells can move on a surface endowed with a gradient of a chemical species with which it has specific affinity. While the above studies were carried out in the biological domain, Bascom[24] et al demonstrated that a drop of oil containing a surfactant can move randomly on a surface. Part of the explanation of this effect, as we know today, is again a differential adhesion of the drop on its advancing and trailing edges due to the disparate adsorption of surfactant on the surface. It was Greenspan[25], who provided the first exposition of the motion of a liquid drop on a heterogeneous surface that is related to the differential adhesion of the drop on its advancing and trailing edges. He predicted on theoretical grounds that a non-wetting drop should move on a surface towards the region of greater adherence and formulated the first hydrodynamic theory underlying such motion, which was further consolidated by Brochard[26] about a decade later. Chaudhury and Whitesides[7] provided the first experimental demonstration of the effect that is described below in some depth. However, before we discuss these results, it is prudent to present an outline of how the gradient surfaces are made that allow such studies to be carried out.

**2.1. Preparation of Gradient Surfaces.** A surface with which to perform the experiments of droplet motion is generally prepared by a diffusion[4,7,27] controlled chemical reaction (figure 1) of a functional alkane with the surface. There is also a beautiful recent study[28] describing the preparation of such a surface using the method of stamping. In all cases, one reactive end (usually a chlorosilane) of such an alkane strongly reacts and/or chemisorbs onto the surface, whereas its alkyl group provides the needed low surface energy or hydrophobicity. The part of the surface where the grafting density of the alkane is highest becomes maximally hydrophobic, whereas the surface becomes progressively hydrophilic towards the other end where the alkane molecules adsorb sparsely. The resultant surfaces can be characterized by measuring contact angles of a probe liquid and, often, with other surface sensitive techniques[7,27] such as ellipsometry and atomic force microscopy. The portion of the surface that exhibits the most linear variation of the surface energy expressed in terms of the cosine of the contact angle is used for the study of droplet motion. A non-wetting liquid placed on such a surface moves toward the region of higher adherence, the underlying physics of which can be understood as follows.

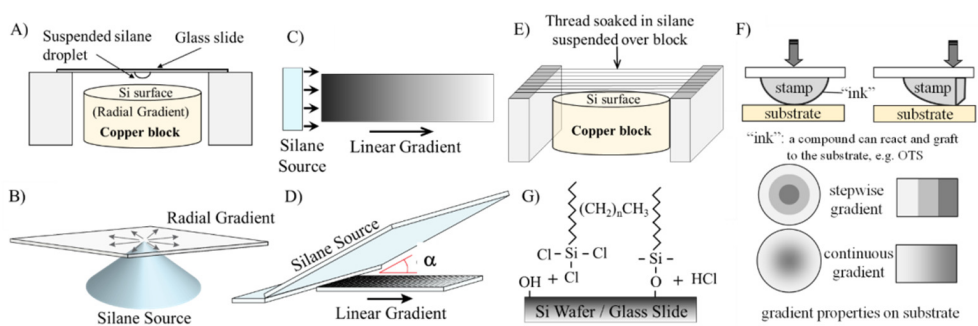

**Figure 1**: A-E) Diffusion controlled silanization of a glass or a silicon wafer produce the gradient surfaces. A radial gradient is prepared (A and B) using either a drop of, or a cone shaped filter paper soaked in, silane. The line source of silane in C) generates a gradient in the form of an error function. D) A filter paper soaked with silane placed at an angle against the substrate produces a longer linear gradient. E) Parallel strings soaked in silane placed above a substrate produce a periodic gradient. F) Schematic of the preparation of gradient using a contact printing technique.[28] G) Schematic of the chemical reaction of alkyl trichlorosilane with the hydroxyl group of a surface. In all cases, the darker region signifies higher grafting density of silane. (B,D Reprinted with permission from (27). Copyright (2012) American Chemical Society)

**2.2. Quantitative Aspects of the Motion of Droplets on a Gradient Surface.** Consider a surface that has a gradient of wettability i.e. $\theta$ varies continuously from one to the other end of a surface. The free energy of adhesion of the drop with the surface at any given location is: $\Delta G = -\gamma_{lv}(1+\cos\theta)\pi R^2$, where, $R$ is the base area of the drop, $\gamma_{lv}$ is its surface tension and $\theta$ is the average of the contact angles on the advancing and the receding edges of the drop on the surface. If the gradient of wettability is continuous along the gradient, the drop experiences an unbalanced Young's force $F_Y = -d(\Delta G)/dx$ [the subscript of $F_Y$ is ascribed to Young] causing it to move towards the more wettable part of the surface accompanied by the dissipation of energy due to viscous motion of the fluid in the bulk and near the three phase contact line. The entropy producing drag force is proportional to the viscosity, radius and the droplet velocity, as in the Stokes equation, i.e. $F_d \sim \eta V R$. Equating $F_Y$ to $F_d$, one obtains the velocity of a drop as:

$$Ca(=V/V^*) \sim R(d\cos\theta/dx) + 2(1+\cos\theta)dR/dx \qquad (1)$$

Equation (1) applies when both $\theta$ [$=\theta(x)$] and $R$ [$=R(x)$] vary along $x$. However, when the base area of the drop does not expand appreciably (i.e. $dR/dx$ is negligible), we have:[14,17,25,26]

$$Ca (= V/V^*) \sim R(d\cos\theta/dx) \qquad (2)$$

where, $V^* = \gamma_{lv}/\eta$ is the capillary velocity. Note that, we avoided writing down a pre-factor to equation 1 or 2, which is usually hard to estimate, although, technically, one may try doing so[25,26,29], especially when the contact angle is less than 90°. Our approach here is to design an experiment to study how Ca varies as $R^*$ [$= R(d\cos\theta/dx)$] and learn from the results something about the nature of this pre-factor. Figure 2 summarizes the results obtained with various liquid drops on the part of a surface that has a linear gradient of wettability that is short and not too steep in order to justify the neglect of the term *dR/dx in equation 1*. In these studies, the contact angle gradient $(d\cos\theta/dx)$ was measured for each liquid drop from the averages of the advancing ($\theta_a$) and the receding angles ($\theta_r$) as a function of distance, which multiplied by the studies performed on different gradient surfaces. These plots are shifted to the origin in the inset. (Reprinted with permission from (17). Copyright (2004) American Chemical Society).

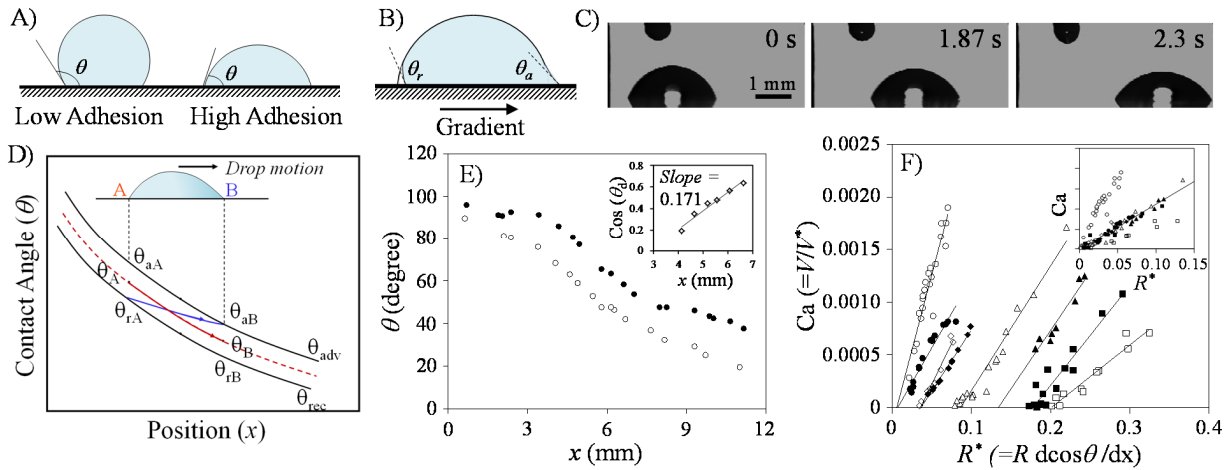

**Figure 2:** A) On a gradient surface, a liquid drop is expected to move from the high to the low contact angle region. B) An exaggerated schematic of a moving liquid drop on a gradient surface. C) Variation of the advancing and receding contact angles of a liquid on a gradient surface is shown schematically. D) Video snapshots of a moving drop of ethylene glycol on a gradient surface. E) Typical values of the advancing and receding contact angle of a water drop on a gradient surface. The inset shows the gradient of the cosine of the average advancing and receding contact angles. F) Capillary numbers of various drops are plotted against the normalized gradient force. Liquids used in this study were dipropylene glycol (○), acetonitrile (●), ethylene glycol (◊,♦), formamide (△,▲) and water (■, □). Different symbols for the same liquid imply

radius of the drop yields a non-dimensional value of the gradient: $R^*$. The measured velocity of a drop is non-dimensionalized by dividing it $V^*$. Several observations of these experiments deserve comments. To begin with, we note that the drops appear to be spherical caps (figure 2C) on such surfaces even when its advancing and receding contact angles are different. What causes this? In order to answer this question, we need to realize that the motion of the drop occurs due to the gradient of Laplace pressure within the drop, which creates the hydrodynamic flow that is balanced by a viscous stress[25], i.e. $-p_x = \eta V_{zz}$ (here and elsewhere, a subscript is used to mean a derivative: $p_x = \partial p / \partial x$; $V_{zz} = \partial^2 V / \partial z^2$). Since the pressure is obtained from the product of the surface tension and the gradient of curvature ($\kappa$) and $V_{zz} \sim \langle V \rangle / \xi^2$ (where $\langle V \rangle$ is the volumetric drop velocity and $\xi$ is the local height of the drop), we have: $\partial \kappa / \partial x \sim Ca / \xi^2$, or

$$\frac{\partial \kappa}{\partial x} \sim \frac{R}{\xi^2} \frac{d \cos \theta}{dx} \qquad (3)$$

Thus for low capillary number flow (or for weak gradient), the curvature gradient is finite, but it is unperceptively small for most part of the drop except very close to the contact line $(\xi << R)$, which is usually not detected within the resolution of a normal optical technique. Another important observation is that the drop of a given liquid has to be larger than a critical size[7,14,17] before it can move. This is, of course, related to contact angle hysteresis that has to be overcome before any motion occurs. Hysteresis[14] modifies equation 2 as follows:

$$Ca (= V/V^*) \sim R(d \cos \theta / dx) - (\cos \theta_r - \cos \theta_a) \qquad (4)$$

Where $\cos \theta_r - \cos \theta_a$ is the hysteresis that is found to increase with $\gamma_{lv}$ (figure 2F) of liquids.

**2.3. Hysteresis.** A drop advances on a surface with a (advancing) contact angle that is larger than that when it recedes (receding angle). A simple generalization of different models[30-34] of wetting hysteresis is that the surface imperfections modify the parabolic energy profile of a drop by creating metastable energy states. The drop can get stuck in any one of these states, but is bound by two extreme states that correspond to the advancing and receding angles. The drop on the surface is therefore like a spring with a spring constant $k_s$ that depends on the surface tension and the contact angle, i.e. $k_s = \gamma_{lv} f(\theta)$ (See Supporting Information (SI) for details). In an ideal (hysteresis free) situation, the radius of the drop deviating from its equilibrium value $R_o$ is readily restored due to this spring force. On a surface with pinning defects, multiple metastable states are created, where the local equilibrium of the drop is determined by the balance of the spring and the pinning forces (See SI for details). From balance of these forces, or from dimensional considerations alone, a zero temperature (i.e. in the absence of a thermal or an athermal activation) description of wetting hysteresis is: $R^{\pm} - R_o \sim \pm(u_o R^{\pm} / k_s \lambda^2)$, where, $R^{\pm}$ is the minimum or the maximum value of the radius of the drop corresponding to the advancing and receding states, $u_o$ is the characteristic pinning energy per defect and $\lambda$ is the average distance between two defects (see SI). The base radius of the drop in an advancing or a receding state can thus be related to its equilibrium value as:

$$R_a = R_o / [1 + (u_o / k_s \lambda^2)] \quad \text{and} \quad R_r = R_o / [1 - (u_o / k_s \lambda^2)] \tag{5}$$

For weak hysteresis case, these two extreme base radii follow a reciprocity rule: $R_a R_r \approx R_o^2$. The physics of disorder deals with a length scale, called the "Larkin Length" or $L$, which, in our problem is $L \sim (k_s \lambda^2 / u_0)^2 \lambda$ (see SI). The physical significance of $L$ is that when it is much larger than the defect correlation length ($\lambda$), the contact line fluctuates sufficiently to probe different defects and the hysteresis decreases, which is the case of weak pinning. On the other extreme, when $L < \lambda$, the contact line is strongly pinned. In our problem, $L/\lambda \sim [R_o /(R_r - R_a)]^2$ is on the order of unity; thus the contact line is moderately pinned on the surface. If we consider that $\gamma_{lv}(1+\cos\theta^{\pm})$ to be the interfacial energy release rate (analogous to the strain energy release rate in fracture mechanics) with an amplification function $\phi$ ($= f(u_o/k_s\lambda^2)$), and $\gamma_{lv}(1+\cos\theta)$ is the thermodynamic work of adhesion, then we may write the following equations for advancing and receding contact angles: $\gamma_{lv}(1+\cos\theta^{\pm})=\gamma_{lv}(1+\cos\theta)\pm\gamma_{lv}(1+\cos\theta)\phi$, or:

$$\cos\theta_r = \cos\theta + (1+\cos\theta)\phi \text{ and } \cos\theta_a = \cos\theta - (1+\cos\theta)\phi \qquad (6)$$

So that for weak hysteresis, $(\cos\theta_a)(\cos\theta_r) \approx \cos^2\theta$; but, $(\cos\theta_a + \cos\theta_r)/2 = \cos\theta$.

This arithmetic rule is what is used above (figure 2D) to estimate the driving force for a drop under a wettability gradient. It should be pointed out that the symmetry expressed in equations 5 and 6 is essentially same as what was proposed by Good[33] nearly sixty years ago in that the cosines of the advancing and receding contact angles are related to the equilibrium angle with a friction factor $\xi(r)$ as $\cos\theta^{\pm} = \cos\theta \pm \xi(r)/\gamma_{lv}$. For strong hysteresis, however, this symmetry is not preserved[34]. This may be so with specific bonds, with which the energy barrier for receding case could be larger than the advancing case leading to an asymmetry of advancing and receding contact angles about the equilibrium value.

In the light of the above general discussion, the trend observed in figure 2F (inset) can be explained if $u_o/\lambda^2$, i.e. the adhesion energy barrier of the liquid with the surface increases more strongly than the surface tension of the liquid. Furthermore, the near independence of hysteresis on the position along the gradient would suggest that $u_o/\lambda^2$ is fairly constant along the gradient. Clearly, more work needs to be done in order to develop a statistical mechanics based model of pinning of three phase contact line at the molecular level, which needs to combine the geometry of the heterogeneity, the interaction strength as well as the molecular volume of the probe liquid, as was observed by Timmons and Zisman[35] to be an important contributor to wetting hysteresis. Another important observation is that all the $Ca$-$R^*$ plots shown in figure 2F (except dipropylene glycol) can be shifted laterally to pass through the origin that nearly superimpose on each other, meaning that the pre-factor of equations 1 and 3 is fairly constant for all the liquids. Although, this observation sketches an apparent success of a hydrodynamic treatment of contact line motion on such surfaces with relatively low hysteresis, the near independence of the pre-factor of the $Ca$-$R^*$ plots contrasts the existing theoretical description[10] of the contact line motion, in which the hydrodynamic drag is inversely proportional to the contact angle. Within the window of the observations reported in the experiments, contact angles varied rather widely: $20^\circ$ to $40^\circ$ for acetonitrile, $40^\circ$ to $57^\circ$ for ethylene glycol, $65^\circ$ to $77^\circ$ for formamide and $63^\circ$ to $90^\circ$ for water. Both the facts that the velocity of the drop is constant along the gradient and is independent of the intrinsic dynamic contact angle of the liquid are, thus, surprising. By the same token, the deviation of the data for dipropylene glycol from the collapse of the rest of the data deserves special mention. Its dynamic hysteresis of contact angle (0.01) expressed in terms of $\cos\theta_r - \cos\theta_a$ also deviates from the value (0.13) measured from the advancing and receding contact angles. For other liquids, the agreement is somewhat better, e.g. the dynamic hysteresis of acetonitrile, ethylene glycol, formamide and water are 0.01, 0.06, 0.21, and 0.27, which can be compared with their directly measured values that are 0.01, 0.1, 0.18 and 0.27, respectively.

**3. MOTION DUE TO EXTERNAL NOISE: VIBRATION.**

**3.1. Blooming Noise* and Activated Depinning**. In order to exemplify how much the structure of interfacial fluid mechanics could possibly differ from that of the bulk, we make a slight digression from our original discussion and point out some interesting features of the motion[36] of liquid drops on a structured (so called superhydrophobic) surface. The physics underlying this discussion, nonetheless, is important when the motion of a drop is induced on a surface endowed with a morphological gradient[37-40]. The results reported in figure 3 are with a PDMS film (polydimethyl siloxane elastomer) decorated with microscopic pillars of uniform cross-section and spacing. When a drop of water is placed on such a surface inclined above a critical angle, it moves down due to gravity. However, at a sub-critical angle of inclination (figure 3), the drop exhibits a critical speeding up dynamics in the sense that it remains stationary for certain amount of time; but then it moves rather suddenly and accelerates[36]. While the drop appears to remain, more or less, in a quiescent state on the surface, microscopic examination of its contact line reveals rich dynamics involving de-pinning events and avalanches (figure 3C). The energy released from such a de-pinning process eventually coalesces and excites the surface of the drop (figure 3E), which we call "blooming noise". When the noise evolves to a sufficient intensity, it helps the contact line to de-pin further, thus the drop begins to move and accelerate on the surface (figure 3D). We believe that this feature of drop motion, in which a growth of fluctuation announces a critical speed up dynamics is in qualitative agreement with a prediction by Pomeau[43] et al. The velocity of the drop, however, does not correlate (figure 3F) in a simple way with its bulk viscosity[36], which indicates it is beyond the description of classical fluid mechanics alone. The scope of these experiments can be extended further by vibrating the surface laterally with a random Gaussian noise and even challenging the drop with a gravitational energy barrier[37] (figure 3G). The data reported in figure 3H clearly show that the dynamics of the drop of either water or glycerol exhibit the classical features of activated barrier crossing. Each of the drops remains arrested in the valley for some time and then it crosses over the barrier quickly with a such as displacement, as an output. Although the strength of "noise" is intimately connected to temperature in a thermal system, it could also be mechanical. When noise is periodic, we call it "structured noise", whereas it is called "stochastic" when it is random. When a system is subjected to an external perturbation, the noise is external. When it is generated by internal mechanical fluctuation, we call it "internal" or a "self-generated noise", similar to a term used in reference 41. In reference 42, a term "mechanical activation" is used, which is also meaningful in some situations.

___________________________________________________________________________________
**\*Note**: In this paper, we frequently use the two terms "fluctuation" and "noise", distinction of which is delicate at times. As far as possible, we reserve the term "noise" in a system as an input and "fluctuation",

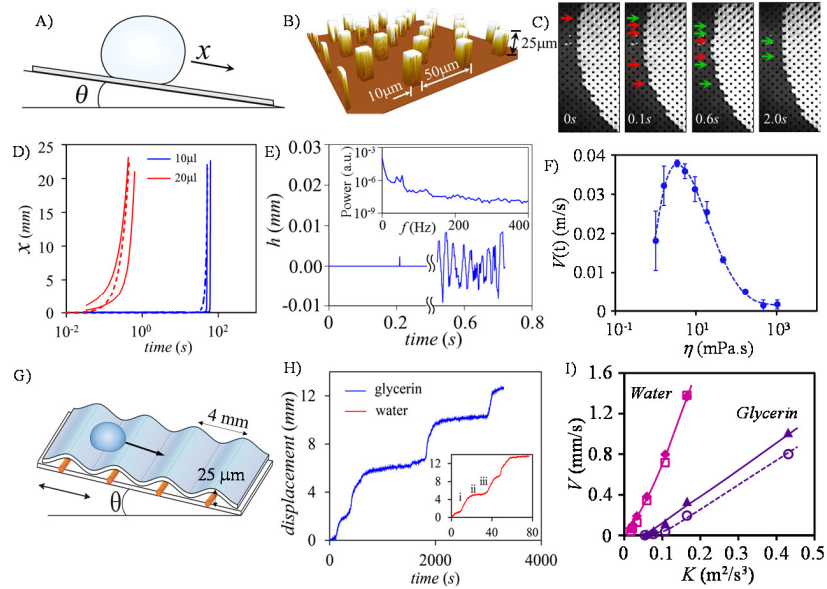

**Figure 3.** A liquid drop (A) on a fibrillar surface (B) exhibits dynamic de-pinning (C) [See Movie 1, SI]. The critical speeding up dynamics (D) is accompanied with surface fluctuation (E) due to the blooming of noise. (F) The average velocity of the drop varies non-trivially with its dynamic viscosity. (G-H) Barrier crossing of a liquid drop on an undulated fibrillar surface exhibits stick-slip behavior. (See Movie 2, SI) (I) Similarity of the average velocity of a drop obtained from barrier crossing (open symbols) and on a flat inclined fibrillar surface (closed symbols). (A-E,H[36] Reprinted with kind permission of The European Physical Journal).

typical frequency that can be sampled from a large numbers of such experiments. When this frequency is multiplied by the spacing between the valleys, we obtain an average drift velocity (figure 3I) that increases with the noise strength [which is defined as: $K=\langle f^2(t) \rangle \tau_c$, $f(t)$ being the value ($m/s^2$) of the random noise pulse and $\tau_c$ is its duration (40 $\mu$s)]. What is intriguing is that these velocities are very close to those obtained for the drop moving continuously on a flat fibrillar surface (figure 3I). This similarity strongly suggests that the motion of the drops on the fibrillated surface is noise activated in the same way as the drops crossing a physical barrier placed on its path. Secondly, the velocity of a drop of glycerol at a given noise strength is only about five times smaller than that of water, even though the difference of their bulk viscosities suggest that this ratio should be of the order of three decades ! All the above observations portray a non-trivial dynamics of motion of a drop on the surface. In general, there are two types of processes controlling the dynamics of a drop. While its bulk motion falls within the scope of classical hydrodynamics, the motion of the contact line involves a pinning/de-pinning transition that shares some common features with the plastic flow involving self-organization of noise[41,42] and activated de-pinning that is beyond the reach of ordinary continuum hydrodynamics. At this juncture, we point out a recent interesting study[44] that reports that the three phase contact line of a colloidal particle at an oil-water interface exhibits a very slow (logarithmic) relaxation with time, which is similar to the ageing kinetics in glassy system. Overall philosophy expressed in that study is similar to what we stated above. For weak pinning, a hydrodynamic description could be used with adequate caution, especially in accounting for the role of a self-generated noise due to de-pinning events and avalanches that could give rise to a dynamic perturbation of the contact line and/or the surface of the drop, itself.

The subject of this paper is how to attenuate hysteresis and/or to use it to our advantage. According to the discussion in the above section, hysteresis is related to the ratio: $u_o/k_s\lambda^2$. A sufficiently strong external perturbation[45-49] to the drop may mitigate the hysteresis, and even convectively pull the contact line out of all the metastable wells. When the noise is stochastic, the contact line may cross the barrier diffusively. These subjects are discussed below.

**3.2. Motion Induced by Symmetric Periodic Vibration and the Roles of the Drop's Eigen Modes.** The effect of vibration on hysteresis can be viewed in the following way. On a surface possessing a gradient of wettability, a liquid drop is effectively subjected to an asymmetric hysteresis as show in figure 2C. When the drop is oscillated, this asymmetry in hysteresis rectifies the imposed inertial force thereby setting it to motion along the gradient.

A sessile drop, however, has two degrees of freedom – one at its center of mass and the other at the center of its base. These two degrees of freedom are coupled by the following equations:

$$\frac{d^2 x_2}{dt^2} + \frac{1}{\tau_B}\frac{dx_2}{dt} + \omega_o^2 x_2 = f(t) \tag{7}$$

$$\frac{1}{\tau_L}\frac{dx_1}{dt} = \omega_o^2 x_2 + \frac{1}{\tau_B}\frac{dx_2}{dt} - \sigma(V)\Delta \tag{8}$$

Here, $\tau_L$ is the Langevin relaxation time (ratio of the mass $m$ of the liquid drop and its kinematic friction coefficient $\zeta$) near the contact line region, $\tau_B$ is the Langevin relaxation time due to viscous friction in the bulk of the liquid, $f(t)$ is the time varying acceleration ($F(t)/m$) with zero mean that the object experiences due to an oscillating force: $F(t)$. $\sigma(V)$ is the signum function of velocity and $\Delta$ is a first order non-linear friction force resulting from the contact angle hysteresis (See also SI for details). The signum function $\sigma(V) = V/|V|$ provides the sign of velocity of the object that takes the values of -1, 0, +1 when $V < 0$, $V = 0$, $V > 0$ respectively thus ensuring that the resistance due to non-linear friction always acts against the motion of the object. Equation 7 has the following (underdamped) solution, which injects a time varying force $A\omega_o^2 \sin(\omega t - \delta)$ to the base of the drop that causes it to move:

$$x_2(t) = \frac{f_o \sin(\omega t - \delta)}{\sqrt{(\omega_o^2 - \omega^2)^2 + (\omega/\tau_B)^2}} = A\sin(\omega t - \delta) \tag{9}$$

$$V_1 / \tau_L = A\omega_o^2 \sin(\omega t - \delta) - \sigma(V_1)\Delta \tag{10}$$

The solution of equation 10, however, is challenging due to the non-linear term $\sigma(V_1)\Delta$. However, we know that the object moves, on the average, linearly when the window of observation time is rather large (at least larger than $\tau_L$). By carrying out an effective linearization[50] of equation 10 in the form $V_1/\tau_L^* = A\omega_o^2 \sin(\omega t - \delta)$ with a difference term: $\varphi = V_1/\tau_L^* - V_1/\tau_L - \sigma(V)\Delta$ and minimizing the expected value of it square with respect to $\tau_L^*$, we have:

$$\frac{1}{\tau_L^*} = \frac{1}{\tau_L} + \frac{\Delta\langle|V_1|\rangle}{\langle V_1^2\rangle} \qquad (11)$$

Both $\langle|V_1|\rangle$ and $\langle V_1^2\rangle$ of equation 11 depend on the deviatoric part of the velocity about the average value. If the base of the drop is driven by an acceleration $A\omega_o^2 \sin(\omega t - \delta)$, its velocity will be of the form $V_1 = -(A\omega_o^2/\omega)\cos(\omega t - \delta_1)$. Using this form of the velocity, the ratio of $\langle |V_1| \rangle$ and $\langle V_1^2 \rangle$ becomes $2\omega/(\pi A \omega_o^2)$. Thus, equation 11 assumes the form: $1/\tau_L^* = 1/\tau_L + 2\omega\Delta/(\pi A \omega_o^2)$. Now, if a force $m\overline{f}$ acts on the drop, a linear response criterion can be used to obtain the drift velocity as:

$$V_d = \overline{f}\tau_L^* = \frac{\overline{f}\tau_L}{\left(1 + 2\Delta\tau_L\omega\sqrt{(\omega_o^2 - \omega^2)^2 + (\omega/\tau_B)^2}/(\pi f_o\omega_o^2)\right)} \quad (12)$$

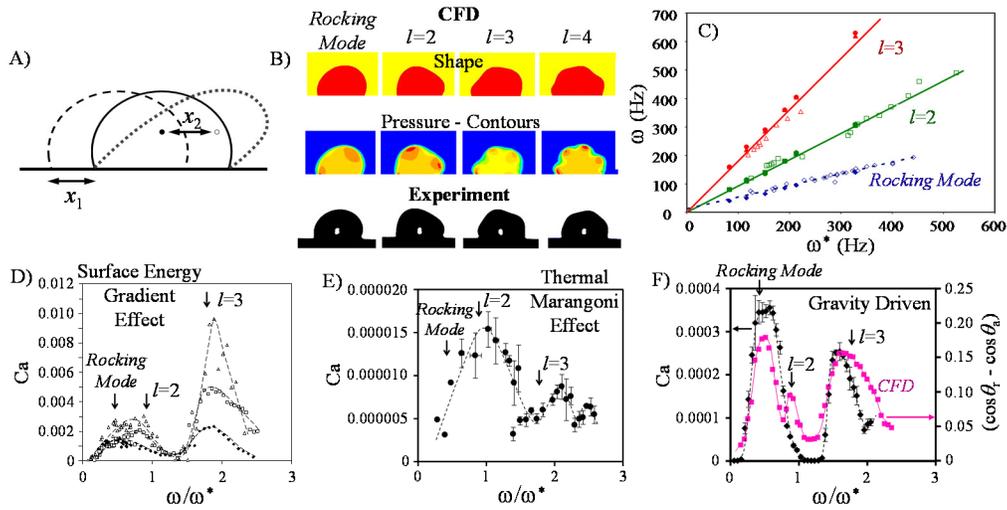

**Figure 4**: A) Schematic of a drop on a laterally vibrating substrate. B) Various spherical modes of a water drop on a surface as obtained from computational fluid dynamic simulation (colored) and those obtained experimentally (black). C) The eigen-frequency ($\omega$) varies with the fundamental frequency of vibration: $\omega^* = (\gamma/m)^{0.5}$, $m$ being the mass of the drop. (D-F) The capillary number of the drop under different biases plotted against the frequency of vibration shows the amplification of velocity at resonance. (D Reprinted with permission from (17). Copyright (2004) American Chemical Society) (F) The frequencies at which maximum amplification (black) occurs are similar to those obtained from CFD simulation (pink). In the latter, the amplification of the difference of the contact angles on the two sides of a drop identifies the resonance frequencies. (B,C,F [51] Reprinted with kind permission of The European Physical Journal).

It is clear from equation 12 that the drift velocity increases with the amplitude of vibration sub-linearly and saturates to $\overline{f}\tau_L$ when $\pi f_o\omega_o^2 \gg 2\Delta\tau_L\omega$. This is when the hysteresis is effectively eliminated. It is also clear that the effect of the hysteresis will be minimum under resonance condition, i.e. when the $\omega \sim \omega_o$. The bias $m\overline{f}$ may be provided by a gradient of wettability, i.e. $\gamma_{lv}(d\cos\theta/dx)\mathcal{A}$, by a gradient of temperature that creates a temperature dependent Marangoni force i.e. $(d\gamma_{lv}/dT)(dT/dx)\mathcal{A}$ (where $\mathcal{A}$ denotes the area under the liquid drop) or, simply, by a gravitational body force ($mg$). In all cases, the maximum velocity is observed when the applied frequency is close to one of the eigen modes of drop vibration, although some differences of the peak positions are noticeable depending upon whether the bias is gravitational, thermal or induced by a gradient of surface energy. Some of these differences are attributed to the differences in contact angles of liquids on a surface. There is also a non-trivial coupling[51] between different modes depending on how the bias acts on the drop. The eigen modes of the vibration of a drop can be precisely estimated[51] by carrying out a numerical solution of Navier Stokes equation (equation 12) with the associated level-set implementation. The eigen modes of the drops obtained from such a numerical method agree favorably with those obtained experimentally. More to the point, the numerically calculated eigen modes of a drop vibrating vertically in contact with the surface are in good agreement[51] with the modes where maximum amplification of velocity occurs.

**3.3. Velocity Amplification and Hysteresis.** From the results summarized in figure 5, it is clear that vibration has strong effects on drop motion with those liquids that exhibit appreciable hysteresis. However, the motions of those drops that exhibit small hysteresis are minimally affected by vibration (Figure 5A). These results can be understood by taking the ratio of the velocity with (eq. 12) and without vibration (eq. 4), the amplification factor (*AF*) becomes:

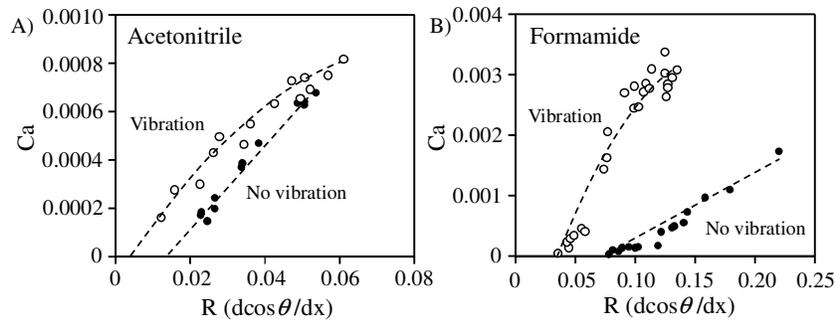

**Figure 5. (A-B)** Effect of a harmonic vibration ($\omega$=100 Hz and $A$=0.3 mm) on the migration of two kinds of liquid droplets on a surface possessing a wettability gradient. Note that acetonitrile (left), which has negligible hysteresis, exhibits smaller amplification of drop velocity in comparison to formamide that exhibits higher hysteresis.

$$AF = \frac{1}{(1-\Delta/\bar{f})\left[1 + 2\Delta\tau_L\omega/(\pi A \omega_o^2)\right]} \qquad (13)$$

According to equation 13, no amplification of velocity occurs when $\Delta$ =0, i.e. *AF* =1. *AF* increases with $\Delta$. Thus, in the presence of vibration, only a small amplification of velocity occurs for acetonitrile, whereas a strong amplification of velocity occurs for formamide (figure 5). There is however one interesting feature of the vibration induced drop velocity that equation 12 does not capture. When the hysteresis is completely mitigated, equation 12 predicts that the drop velocity should be $\bar{f}\tau_L$, which can be read out from the master curve obtained by shifting all the Ca-$R^*$ plots of figure 2F passing through the origin. The velocities (figure 4D) of some of the liquids at the resonance conditions are, however, frequently larger than $\bar{f}\tau_L$. The failure to capture this interesting feature of vibration induced droplet motion is that a third degree of freedom arising from the vertical component of shape fluctuation has not been taken into account in equation 12. The non-linear coupling of the vertical and horizontal vibration can give rise to additional inertial force. Recent studies[52,53] found that such a coupling is enough to break the symmetry of the periodic force that a drop experiences on a homogeneous surface without any additional bias. In our own study, we noticed that vertical shape fluctuation resulting from the horizontal vibration of a substrate can sometimes be so severe that the drop can even detach from the surface[54] (see below). On a strongly hysteretic surface, the vertical fluctuation is less significant, while it is more pronounced on a lower hysteretic surface. The differences of the behavior of formamide and water may result from the differences of hysteresis of these two surfaces, in that the former liquid exhibits lower hysteresis than water (figure 4D). The intermediate behavior of ethylene glycol could be because of its higher viscosity that damps the vertical shape fluctuation more than the other two liquids.

**3.4. Drop Motion Due to Shape Fluctuation (Stretching and Compressing the Droplet).** Here we report how hysteresis rectifies the repeated stretching and squeezing a drop between two nearly identical gradient surfaces, which was first discovered by Daniel[17] et al, and followed up in reference 27 (See Movie 3, SI). The periodic squeezing and extension of a drop between two plates[17,27] cause the advancing and receding contact lines to be pinned and de-pinned periodically (figure 6). The drop thus undergoes an inchworm like motion with a speed being the product of $\Delta x$, net displacement of the drop in a complete cycle of oscillation, and the driving frequency ($\omega$) (figure 6). In this mode of motion, hysteresis does not appear[27] explicitly in the equation of motion as long as it gives rise to certain amount of rectification per cycle. After Daniel[17] et al,

Prakash[55] et al found that the asymmetry needed to generate such a directed motion can also be created by placing the drop between two flat slanted surfaces; the physics being the same.

**3.5. Asymmetric Structured Noise.** The above example illustrates that hysteresis manifesting in terms of a non-linear friction is quite effective in rectifying a symmetric periodic vibration that, in turn, gives rise to directed motion of the drop in the presence of a net bias acting upon it. In this section, we illustrate another method of producing directed motion that does not require any external bias[51,54,56] at all. Consider an asymmetric vibration[54] of the plate supporting a drop is of the type: $f(t) = A_o[\cos(2\pi\omega t) + 2\cos(4\pi\omega t)]$ exhibiting an asymmetry of amplitude. The drop, thereby, experiences an acceleration – $f(t)$. If only a linear friction operates at the interface, the drop does not move because the temporal averages of all terms on the right hand side of equation 7 are zero. A non-linearity is needed to break the symmetry of time or displacement for the drop to reach its original position after going forward or backward. This can be achieved with a non-linear friction $\pm \Delta$ under the condition that $|f(t)| > |\Delta|$. With a finite $\Delta$, the average velocity is $\langle V_1 \rangle = -\Delta \tau_L \langle \sigma(V) \rangle$, which is non-zero; hence the drop experiences a net effective inertial forcing causing a hysteresis of displacement in each cycle of acceleration. The effect of hysteresis on drop motion can be illustrated with three types of surfaces: one with negligible hysteresis, the second with small but finite hysteresis and the third with a large hysteresis. Figure 7B illustrates that a drop does not exhibit any motion either with a negligible or with a large hysteresis (See Movie 5, SI), but it executes a net directed motion with a moderate hysteresis. The drift motion of such a drop can be described quite adequately by solving numerically two coupled equations of motions (equations 7 and 8): one for the center of the mass of the drop and the other for the center of the base and by replacing the $\sigma(V)$ with a hyperbolic tangent function [tanh($\beta V$)], which behaves like a signum function with a reasonably high value of $\beta$. Solutions of these two coupled equations quantitatively predict[54] that no motion should occur for $\Delta=0$ or for $\Delta>>0$, but for a moderate value of $\Delta$ (figure 7B and 7D). The solution of these equations also predicts that the in-phase and the out-of-phase motions of the two degrees of freedom of a drop on a surface can also lead to a flow reversal. This effect is dramatized in figure 7 where a large drop moves upward, but a small drop moves downward along gravity. For a drop with a given size, the flow reversal also depends on the vibration frequency, the physics of which can be captured using the tools of computational fluid dynamics[51].

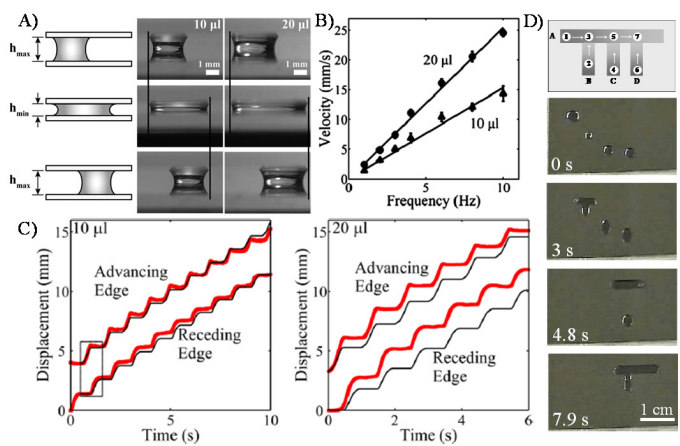

**Figure 6**. Squeezing and stretching of a drop (A) confined between two gradient surfaces lead to a net motion of the drop, which increases with the frequency of vibration (B). The inchworm motion of the advancing and receding edges of two drops are shown in (C). The thicker (red) lines are experimental and the thinner (black) lines are theoretical. (A-C Reprinted with permission from (27). Copyright (2012) American Chemical Society) (D) Schematic of a drop fluidic reactor and the video-microscopic images of the sequences of the motion of drops and their fusions. (See Movie 4, SI)   (Reprinted with permission from (17). Copyright (2004) American Chemical Society)

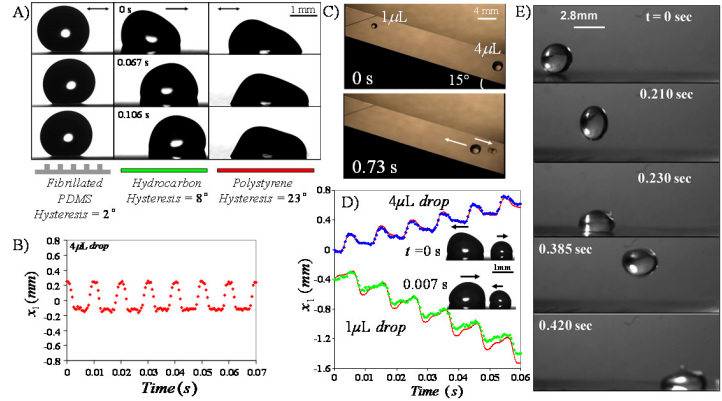

**Figure 7.** The motion of water drops (4 $\mu$l) induced by an asymmetric vibration[54] (figure A) of the form: $A_o[\cos(2\pi\omega t) + 2\cos(4\pi\omega t)]$ ($A_o = 33 m/s^2$) depends on hysteresis. High speed tracking of the drop on zero hysteresis surface is shown in (B) and that with medium hysteresis is shown in (D). (C) Flow reversal of two different size drops (4 $\mu$l and 1 $\mu$l) (See Movie 6, SI). In (D) blue and green dotted curves are experimental, whereas the red continuous curves are from simulation. In all the above experiments, (E) A 10 $\mu$l drop on a fibrillated PDMS surface moves with a strong vibration ($A_o$= 90 $m/s^2$) by detaching from and reattaching to the surface. (A-E Reprinted with permission from (54). Copyright (2011) American Chemical Society)

Even though a small drop of water vibrated on a hysteresis free surface does not exhibit any drift, it does so when a larger drop is vibrated vigorously. In this case, the shape fluctuation in the vertical direction is so intense that the drop is detached from the substrate. The drop flies in the air, re-deposits on the surface and the motion continues (figure 7E). We believe that the symmetry breaking non-linearity here comes from the coupling of horizontal and vertical shape fluctuations of the drop, the underlying physics of which is, perhaps, similar to that observed by Brunet and others[52,53]. Vibration induced detachment of drop has previously been observed[57], but in that case, detachment occurred vertically along the direction of vibration.

**3.6. Potential Applications.** Controlled motion of a liquid drop on a surface induced by external vibration is of potential importance in various technological settings, especially in conjunction with another type of gradient designed upon it. Figure 6D illustrates how such a guided motion can be developed on a surface having a patterned wettability gradient channels[17]. More interesting motions can also be created by an asymmetric vibration, in conjunction with a thermal gradient, which is of particular interest to chemists and engineers as the rates of various chemical reactions or phase transformations are thermally induced. Let us take a simple case of the solubility of a surfactant in water. It is known that various surfactants form micellar solutions at room temperature; however, when heated to higher temperature, the surfactants precipitate thus turning the solution translucent. This transition, called the cloud point transition, can be easily observed[56] by driving a liquid drop from the colder to the hotter region of a surface having a temperature gradient that also allows precise identification of the cloud point temperature (figure 8A) (See Movie 7, SI). Now, as a drop moves from the hotter to the colder part of the surface, it moves faster as the Marangoni force adds to the inertial force than when it moves against the thermal gradient. By appropriate control of the speed of the drop, the underlying hysteresis of the micellar transition can also be studied. The technique is also poised for more advanced processes. For example, drops can be cycled in a loop, fused, stopped, and then switched to motion[56] (figure 8) again using computer assisted commands that may have potential applications in various biochemical reaction processes.

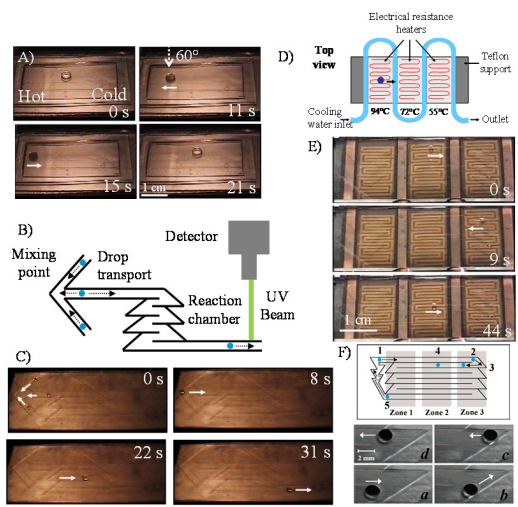

**Figure 8**: A) A drop of 1 % Triton-X solution in water moves across a temperature gradient. The drop turns translucent at the cloud point temperature. The drop can also be propelled backwards where thermal Marangoni force assists its motion. B) Schematic of a fluidic device that allows mixing of three different drops and then transport it through a reaction chamber. C) Video snapshots of drop fusion processes. D) Schematic of a fluidic device having different temperature zones isolated by cooling pipes. (E) Video snapshots of a drop moving back and forth on such a device. F) The motion of a drop can be guided by the contour of a channel. (A,B,E,F Reprinted with permission from (56). Copyright (2005) American Chemical Society)

**3.7. Random Noise, Contact Angle Fluctuation, Relaxation and Driven Diffusion.** Let us consider a linear Langevin equation of an object, where the motion is driven by a bias and a Gaussian noise, but controlled by a linear kinematic friction: $dV/dt + V/\tau = \bar{f} + f(t)$. The solution of this equation gives the average velocity as: $\langle V \rangle = \bar{f}\tau$, which is exactly the same as that in a noise free situation. Thus a linear system gains nothing from noise as far as drift velocity is concerned (see also equation 13). If, however, there is a non-linearity governing the motion, then the drift velocity is enhanced. This was evident in the spreading of a thin liquid film governed by a non-linear equation of motion as was demonstrated in recent reports[58,59] that showed that thermal noise can enhance both of their wetting[58] and de-wetting rates[59] on a surface. The Langevin equation can also become non-linear due to a non-linear friction, as was showed in the context of stochastic sliding by Caughey and Dienes[50] and later by de Gennes[62]. These authors[50,62] argued that an object pinned on a surface due to Coulombic dry friction should exhibit drifted diffusive motion in the presence of a random noise, with the drift velocity being strongly dependent on the strength of the noise.

The problem of a drop pinned on a surface due to defects is similar to that of stochastic sliding[50,60]. To begin with, we show in figure 9A that a sessile drop of water subjected to a random Gaussian noise exhibits vigorous fluctuation, and reaches the same final state starting from either an advanced or a receding state, often exhibiting stick- slip behavior. The magnitude of this fluctuation is inversely proportional to hysteresis (figure 9B) and the probability density of the contact line displacement fluctuation (figure 9D) is somewhat non-Gaussian (see SI).

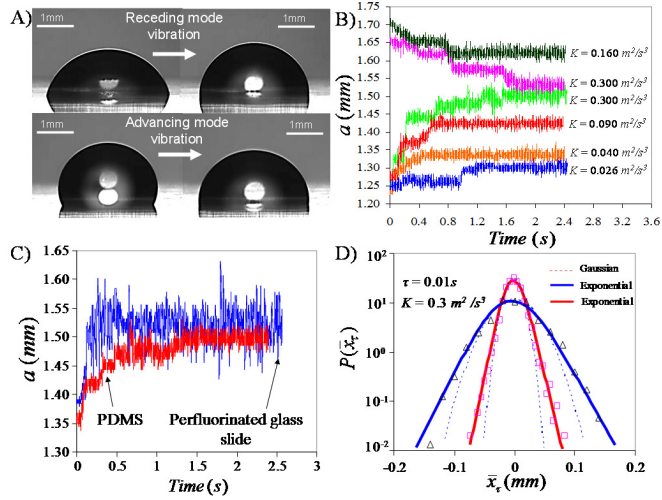

**Figure 9**. (A) Vertical vibration of a drop of water on a glass (or silicon wafer) with a high strength random Gaussian noise induces it to reach the most stable state. (B) The relaxation kinetics of a 10 µl drop of water depends on the noise strength. (C) Fluctuation of the drop increases as the contact angle hysteresis (CAH) decreases. CAH=24° on a per-fluorinated glass slide, CAH= 45° on PDMS (Sylgard-184) elastomer. (D) The probability density function of the displacement fluctuation is somewhat non-Gaussian indicating that hysteresis is not eliminated. (Reprinted with permission from (61). Copyright (2010) American Chemical Society)

This indicates that hysteresis, by itself, does not have to be eliminated as the contact line can escape from the metastable states diffusively (see below for more evidence). We now consider what happens when the drop is subjected to both a random noise $f(t)$ and a bias[60-62] (figure 10). Understanding the dynamics of the contact line here requires solutions of two coupled equations (equations 7 and 8). However, here the random forcing term (via $x_2(t)$) has a stochastic as well a periodic component as is evident in the time series and the power spectral density of the fluctuation of the surface of the drop (figure 11). Nevertheless, the fluctuation of the effective acceleration $\omega_o^2 x_2(t)$ leads to a diffusive motion of the base of the drop (10 µL) on a gradient surface. ($d\cos\theta/dx = 88\,m^{-1}$) (Reprinted from (20) with permission from Elsevier) .C) Random fluctuation of a water drop (10 µL) on a hydrocarbon treated glass surface that underwent random lateral vibration at noise strength 0.005 m$^2$/s$^3$. D) FFT of the fluctuation data of the drop in C yields the noisy power spectrum showing the resonant modes.

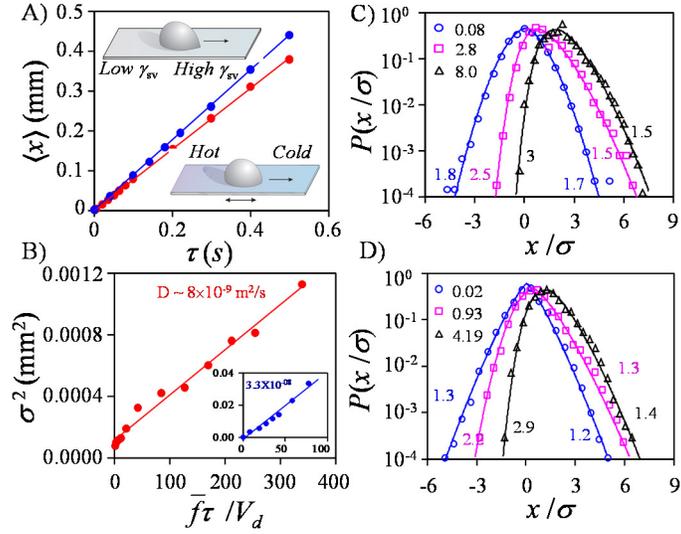

**Figure 10.** The drift (A) and the variance (B) of the displacement fluctuation of a drop biased either by a wettability ( 10 $\mu$l drop, $d\cos\theta/dx = 88\,m^{-1}$ ) or a thermal gradient (4 $\mu$l drop, d$T$/d$x$ = 0.45 °C/mm) obtained with vibrations of noise strengths of 0.02 m$^2$/s$^3$ and 0.8 m$^2$/s$^3$, respectively. The corresponding pdfs of the displacement fluctuations are shown in figures C (surface energy gradient) and D (thermal gradient). The time scales (s) are given in the upper left. At short time, the pdf's are non-Gaussian, but it becomes skewed Gaussian at long times. The left or the right branch of the displacement fluctuation can be fitted with a stretched Gaussian function, the exponent of which is shown next to each pdf. In figure B, the time is non-dimensionalized as $\overline{f}\tau/V_d$. The bias for the wettability gradient is estimated from the surface tension of the liquid and the gradient of wettability that for the thermal gradient was estimated be inclining the surface upward till the drop stops moving that yielded $\overline{f}$ ~ 0.11 $m/s^2$.

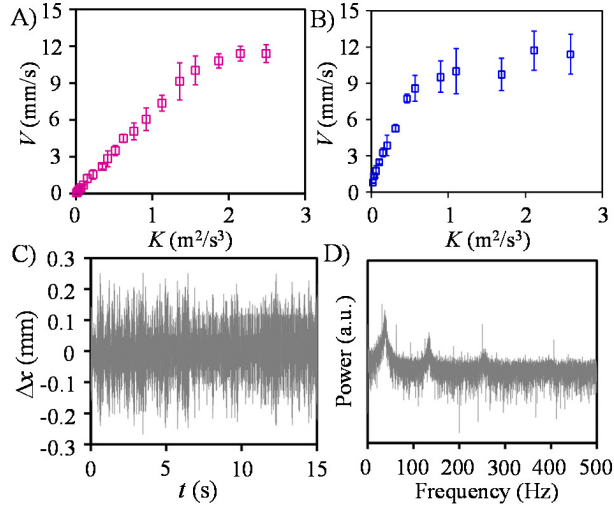

**Figure 11**: A) Drift velocity of a water drop (8 µL) on a fibrillated silicone surface inclined at 2° as a function of the strength of a Gaussian vibration of the substrate. B) Similar data for a water

drop irrespective of whether the bias is applied by means of a gradient of surface energy or a gradient of temperature (figure 10). Both the average drift $\langle x \rangle$ and the variance of the displacement $\langle x^2 \rangle$ are linear in time, from which a drift velocity and a diffusivity of the drop can be estimated, which can even be used (if desired) to estimate an effective temperature in each case using the Einstein's ratio of diffusivity and mobility: $D/\mu$. We addressed this issue in a previous publication[63], where we demonstrated that $T_{eff}$ obtained from $D/\mu$ agrees rather well with that estimated from other methods, i.e. a fluctuation theorem or a barrier crossing experiment. Here, we do not wish to go into these types of details, except reminding the readers the key attributes of a barrier crossing experiment summarized in figure 3, which are paradigmatic to hysteresis. If the motion of a drop on a flat inclined fibrillar surface (as was reported in figure 3) is carried out at higher noise strengths, the drift velocity ultimately saturates (figure 11A) as the non-linear friction is completely mitigated and the dynamics is controlled by a linear friction. A similar behavior is indeed observed with a drop on an unstructured surface, where the bias is due to the gradient of adhesion energy (figure 11B). These results can be captured rather nicely by defining two relaxation times of the drop: one is the classical Langevin relaxation time ($\tau_L$) and the other being the noise strength dependent effective relaxation time[50,63] ($K/\Delta^2$), i.e. $1/\tau_L^* = 1/\tau_L + \Delta^2/K$. Using a linear response relationship between the drift velocity ($V_d$) and the applied force (in this case, force per unit mass), i.e. $V_d = \bar{f}\tau^*$, we can write[63]:

$$V_d = \bar{f}\tau^* = \bar{f}\tau_L/(1+\Delta^2\tau_L/K) \tag{14}$$

Equation 14 suggests that $V_d$ should increase sub-linearly with $K$ at low noise strengths, but it should saturate to $\overline{f}\tau_L$ at high noise strength, as is observed experimentally (figure 11B). At this juncture, we point out an amusing, albeit distant, similarity between noise strength dependent drift velocity of these drops and the well-known behavior of electronic conductivity ($\sigma$) in non-crystalline solids[64] that, in varieties of situations, varies with temperature as $\ln\sigma \sim -T^{-0.25}$. In reference 36, we pointed out that the noise strength dependent drift velocities of various liquids can also be expressed in terms of a sub-Arrhenius form: $\ln V_d \sim -K^{-0.2}$.

## 4. MOTION DUE TO SELF-GENERATED NOISE.

**4.1. Driven Diffusive Motion of Drop Due to Self-Generated Noise: Drop Coalescence.** The above examples are useful expositions of how a mechanically induced noise diffusively de-pins a contact line from the metastable states that is advantageous to generating driven diffusive motion of a drop. Motion induced in this way, however, is of limited use as it is not always possible to subject a drop to external vibration. What would significantly extend the scope of the idea is if the noise is self-generated, which is possible at sub-micrometer length scales via thermal fluctuation[65] that, however, is not effective at meso-scale levels. In this section, we would like to establish that the noise generated by coalescence of droplets could provide the needed impetus for such types of motions. To motivate the discussion, let us consider the case of the wetting of a small drop of oil (e.g. hexadecane) deposited on a rigid flat substrate, which spreads extremely slowly because of the huge hydrodynamic resistance at the three phase contact line and in the precursor film regime. A simple solution to the problem is to cover the surface with condensed small drops of an immiscible liquid, such as water, and then allow the oil droplet to spread over it. Here the driving force is enhanced due to the additional roughness produced by the condensed drops which can also be accomplished by a rough surface or a surface covered with small particles. Spreading of oil over water droplet covered surface is unique in that the water droplets (i.e. the pinning sites) are dislodged[66] from the solid substrate by an attractive capillary force

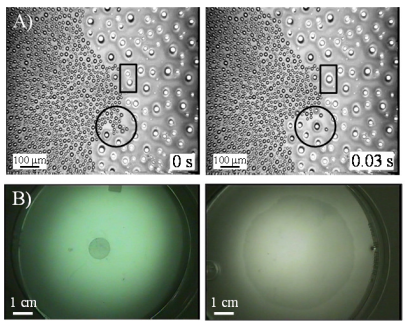

**Figure 12**: A) Sequence shows spreading of a hexadecane drop over condensed water droplets on polystyrene surface. Coalescence of the water drops are highlighted by rectangles and circles. B) A drop of oil (octane, 30 µL) containing 10% crosslinkable polydimethylsiloxane spreads much more over polystyrene covered with condensed droplets (right) than on plain polystyrene (left). After octane and water droplets evaporate, a thin PDMS film (1.5 µm thick and 5.4 cm diameter) is formed (right).[66] (Reproduced by permission of The Royal Society of Chemistry)

(mediated by the thin oil film), which eventually coalesce (figure 12). As the excess surface energy of coalescence is converted to vibration, it induces a large fluctuation in the system thereby enhancing the rate of spreading. A practical example of this effect is given in figure 12 that shows how a thin polymer film can be produced using this method without the need for spin coating. Coalescence of drops can also be used to tackle problems involving water and thermal management technologies. For example, let us consider the case of a heat pipe, which is a sealed tube filled with a small amount of liquid in contact with its vapor as shown in figure 13A. One end of the tube is exposed to a high temperature source so that part of the liquid evaporates. The other end of the tube is in contact with a heat sink so that the vapor condenses and sticks to the surface of the tube as droplets. How wonderful it would be if the liquid droplets flowed back to its original source so that the evaporation-condensation cycle could be repeated, thus siphoning the heat continually from the source to the sink. Although wicking is a widely used method to achieve this objective, it might also be possible to bring the liquid drops back to the source by designing a surface energy gradient on the inner surface. However, hysteresis still prevents the small drops from flowing back to the source. An externally generated vibrational noise can indeed assist the backflow of the drops as demonstrated with some examples in figure 13B-13D. What is more important in the context of the present discussion is that when two drops coalesce, the fused drop moves faster than the parent drops. This principle is explained with an experiment, in which two drops, one small and the other large, are placed on a surface having a gradient in wettability. Here, the larger drop moves towards and coalesces with the smaller drop. As the immediate contact angle following coalescence is less than both the advancing and the receding contact angles, the drop retracts from the trailing side, while the frontal side remains pinned (this is indeed similar to that shown in figure 6). This leads to a rather fast motion of the larger drop towards the smaller drop, which is completely different from what happens with coalescence on a homogeneous surface, where the smaller drop moves towards the larger one.

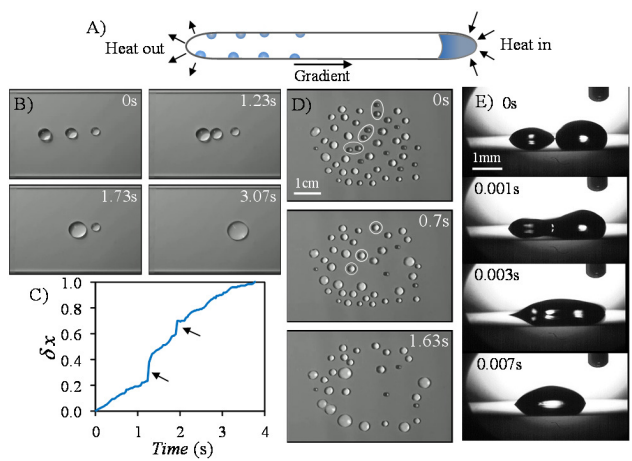

**Figure 13**: A) Schematic of a heat pipe showing the cycle of water evaporation and condensation leading to thermal siphoning. B) When an external random noise is provided to a surface having three drops along a line, they sequentially coalesce to form a larger drop showing enhancement of the speed (arrows in C) when coalescence occurs. Here the travelled distance is a fraction of total distance (See Movie 8, SI)  D) randomly vibrated drops coalesce on a surface having a radial gradient, and move towards the outer hydrophilic region of the surface. E) Coalescence effectively shifts the center of mass of two fused drops of base radii 0.9 mm and 0.8 mm respectively towards the region of higher wettability.

This phenomenon, where the motion of a fused drop is biased along the gradient, is accentuated with the coalescence[4] of multiple drops that happens with a vapor condensing on a surface. In order to demonstrate this principle, let us consider a horizontal metal disc, the lower part of which is convectively cooled, whereas its upper surface that is endowed with a radial gradient of wettability is exposed to a flowing steam. The control parameter here is the degree of subcooling, $\Delta T$ that is the difference between the temperature of the steam and that of the surface. This $\Delta T$ controls the rate of condensation of the water vapor on the surface, the growth rate of the droplets and the rate of their random coalescence. During each coalescence event, the fused drop advances towards the more wettable part of the gradient with its velocity being proportional to the rate of condensation (figure 14B). At high rate of condensation, the fused drops are found to move with a speed reaching about 1 m/s. As the latent heat of condensation is effectively removed by conduction through the metal block, we have the condition for steady state heat flux: $J_q = \dot{m} H_m = k_m dT/dx$, $\dot{m}$ being the condensate flow rate, $H_m$ is the latent heat of vaporization of water per unit mass, $k_m$ is the thermal conductivity of the metal block and $dT/dx$ is the temperature gradient along the axis of the block. Since, $\Delta T$ controls the rate of condensation, we expect that the heat flux $J_q$ to be proportional to it. The results summarized in figure 14D indeed show that although $J_q$ increases with $\Delta T$ on a gradient surface much more than that on a uniformly hydrophilic surface; it ultimately reaches a constant value. Let us also note the similarity of the behavior of $J_q$ (and thus $\dot{m}$) versus $\Delta T$ and that of the drift velocity ($V_d$) of a drop on a surface as a function on an external noise (figure 11B) strength. We express the noise strength[20] to be: $K = (\gamma_w k_w \Delta T / m H_v)$, which is the power generated per unit mass of liquid per unit time due to coalescence. Here $\gamma_w, k_w$ and $H_v$ are the surface tension of water, the thermal conductivity, and the latent heat of vaporization of water per unit volume; $m$ is the mass of an average sized water drop of density $\rho$ and radius $R$. The physical significance of the above definition of noise strength is that both the thermal conductivity of water ($k_w$) and the degree of subcooling ($\Delta T$) enhance the growth rate of the drop and thus the energy output resulting from coalescence. As the latent heat ($H_v$) increases, smaller rate of condensation is needed for a given heat flux. Since, we can also express the bias as: $\bar{f} \sim (\gamma_w / \rho R)(d\cos\theta / dx)$ we obtain an equation for the heat flux[20] in terms of $\Delta T$ as follow: $J_Q \sim B_1/(1+B_2/\Delta T)$ meaning that the effect of hysteresis could be mitigated completely so that the heat flux would be nearly independent of $\Delta T$ at rather high value of sub-cooling.

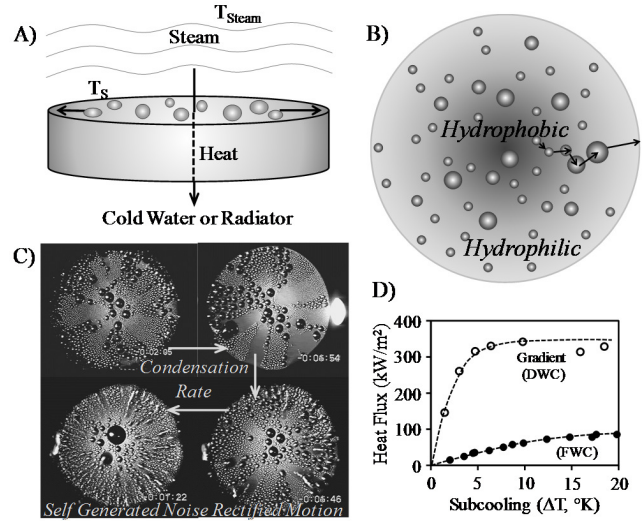

**Figure 14**: A) Schematic of a heat exchanger showing condensation of steam as droplets on a conducting surface that is continually cooled from the bottom. B) The condensed drops exhibit random motion followed by coalescence and continue to move towards it periphery C) Video snapshots showing water drops are removed more rapidly as the condensation rate increases on a radial gradient surface (1 cm diameter) (See Movie 9, SI) D) Heat flux aided by surface energy gradient (DWC) is much larger than that obtained with film-wise condensation (FWC)

**4.2. Diffusive Motion of Droplets: Random Motion with Absorbing Boundary.** It has already been established that coalescence of drops leads to a self-diffusive motion, thanks to the early study of Martin[19] et al. By recognizing this fact, we show below that a bias is not an absolute requirement for generating directed motion of droplets on a surface when coalescence occurs at a rapid rate so that its flux can be totally diffusive[20] as is the case with the random walk of a Brownian particle with an absorbing boundary. This concept is illustrated with a surface (figure 15) made up of alternate hydrophobic and hydrophilic stripes. The observation that the contact angle of a small drop of water is independent of its position on the hydrophobic stripe testifies to the fact that there is no gradient of wettability[20]. In an experiment similar to that described above, vapor condenses as a thin film on the hydrophilic stripe, but as droplets on the hydrophobic stripe that undergo a self-avoiding random walk after coalescing with each other. Close examination of one such drop illustrates the sequence of events. A drop moves only when it coalesces with another drop, in the absence of which the drop remains pinned on the substrate. However, as soon as a coalesced drop reaches the hydrophilic/hydrophobic boundary, it is immediately removed by absorption, which helps continuation of a net diffusive flux of the drops from the hydrophobic to hydrophilic part of the striped surface[20]. We point out here that a similar motion of coalesced droplets has also been observed with a mist of microdroplets impinged either on a surface having a wettability gradient[17] or on a surface having alternate stripes of hydrophobic/hydrophilic zones[20] (See Movie 10, SI). A complete description of such a diffusive motion of coalesced drops has to wait for a full statistical mechanical analysis of the problem. Nevertheless, we expect that such a diffusion will be non-classical in that the velocity distribution would be non-Gaussian for two reasons: a non-linear friction operates at the interface and the noise pulses have a certain pause time following an exponential or a Poisson distribution. However, in analogy to the diffusive flux of a species, the diffusive current of the coalesced drops from the hydrophobic to hydrophilic stripes should be of the form: $J = D\Sigma(L_1/L_2)$, where $D$ is the diffusivity and $\Sigma$ is density of drops per unit area near the middle of the hydrophobic stripe, $L_1$ and $L_2$ being its width and length respectively. Since[62], $D \sim K\tau^{*2}$, and $K = (\gamma_w k_w \Delta T / mH_v)$ (see above), we present (tentatively) an expression[20] for the diffusivity of the coalesced drops as follows:

$$D \sim \frac{\gamma_w k_w \Delta T}{mH_v} \left[ \frac{\tau_L}{1 + (m\Delta^2 \tau_L H_v / \gamma_w k_w \Delta T)} \right]^2 \qquad (15)$$

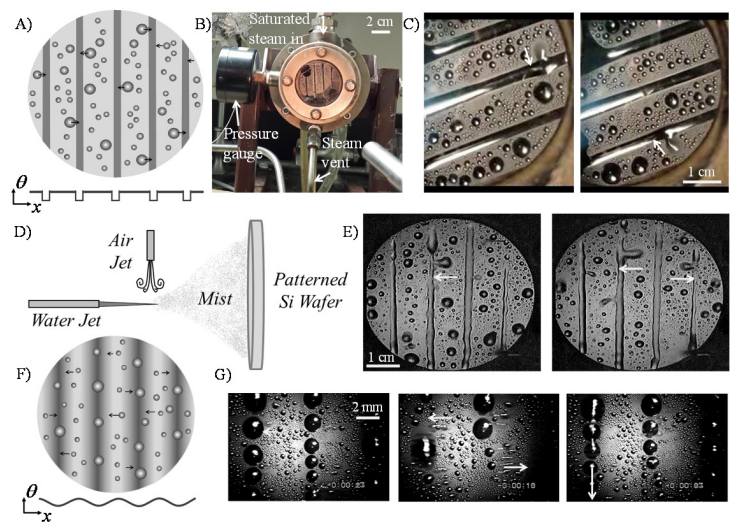

**Figure 15**: A) Schematic of the surface that has alternate hydrophobic (9 mm width) and hydrophilic (3 mm width) stripes. This surface is used in experiments described in B, C, D and E. (Reprinted from (20) with permission from Elsevier) B) A heat exchanger showing drop-wise condensation on the hydrophobic stripe, but film-wise condensation on the hydrophilic stripe. C) Sequence of video images show that the condensed drops continually move towards the hydrophilic stripes. D) Schematic of the impingement of mist of water on the striped surface. E) Video snapshots showing the coalescence induced motion of the drops (originating from the mist) from the hydrophobic to hydrophilic stripe of the surface. (See Movie 12, SI) F) Schematic of the surface with a periodic gradient. G) Video snapshots of the lateral motion of drops on a periodic gradient surface. In all the above, arrows indicate the direction of motion of drops.

An extension of the above striped pattern on a surface is a sinusoidally varying gradient of wettability that can be produced according to a method described[4] in figure 1E (See Movie 11, SI). The drops condensing on gradient part of the surface move laterally and rapidly to the hydrophilic part. When such a surface is placed vertically, the drops collected on the hydrophilic area can be drained down by wicking and or gravity. Significant enhancement of heat transfer can be achieved using either of the two methods[4,20].

## 5. SUMMARY AND PERSPECTIVE.

We discussed how drops can be propelled on a surface due to a wettability gradient and how hysteresis impedes such a motion. We then showed how hysteresis can be used to rectify a periodic vibration like a diode does to a periodic voltage fluctuation. With a Gaussian random noise, a driven diffusive motion of the drops can be set up on a surface, which can also be induced by a self-generated noise due to coalescence. Finally, we pointed out that a continuous gradient of a potential is also not needed to generate droplet motion, if the randomly moving drops are removed by absorption that maintains a constant diffusive flux of droplets from the hydrophobic to hydrophilic part of a surface. This type of motion of drops increases the rate of condensation of vapor on a surface, thus enhancing the efficacy of heat exchanges and heat pipes. One of the main points of this article is the importance of defects that pins the contact line by projecting metastable energy states. Consequently, a drop may not move on a surface even though there could exist a gradient of surface potential to induce such a motion. The pinning of the contact line and its release from the metastable states clearly give rise to rich dynamics that is not captured by a continuum level hydrodynamics. We expect that the dynamics involving strong pinning/de-pinning events to be more amenable to the physics underlying plastic flows. For weak pinning, a hydrodynamic description might be used with adequate caution, especially in accounting for the role of a self-generated noise due to de-pinning events and avalanches that could give rise to a dynamic perturbation of the contact line and/or the surface of the drop.

Hysteresis, as a first order non-linear friction[50], is one way to break the symmetry of a periodic vibration, the physics of which has some semblance to what has been suggested by Mogilner[67] et al to explain the motion of protein motors. In that case, a similar non-linear friction is thought to rectify the asymmetric velocity fluctuations associated with the changes of protein's conformation. Other non-linearities may also develop due to the coupling of the horizontal and vertical shape fluctuations of the drop as was observed first by Brunet[52] et al. Perhaps, a richer dynamics of the drop may be discovered by carefully designing experiments in which these non-linearities interact with each other co-operatively. A point that deserves careful consideration in future is how exactly the external noise mitigates hysteresis. Equation 14 captures the diffusive displacement of the drop that enables it to probe various metastable states. However, if the contact line itself could fluctuate laterally, then the term $u_o/k_s\lambda^2$ [equation 5] that is the source of hysteresis has to be multiplied by the fraction of the active pinning sites that decreases with noise as: $1/[1-\exp(-u_o/T_{eff})]$. Hysteresis would thus decrease and the activated Larkin length could also diverge above certain $T_{eff}$. Although a drifted diffusive motion can be generated by an external noise, a random noise can also develop when multiple drops coalesce with each other. This system then is closer to, but still not same as, a thermal system in that the noise is self-generated, while no fluctuation-dissipation type relationship exists between the noise and the friction. From the viewpoint of non-equilibrium thermodynamics, it would be prudent to focus on how many different ways noise can be self-generated, the practical pay-off of which is unbounded. For example, here is an interesting possibility (M. Grunze, personal discussion) to consider. The water molecules could physisorb and grow to a very thin condensed film on a hydrophobic surface that must break up at some critical thickness, which could result from a combinations of reasons, i.e. hydrophobic attraction between air-vapor and water-substrate interface, thermal Marangoni flow on a perturbed surface, long range van der Waals attraction and nucleation of any dissolved vapor bubbles (see SI). All these could lead to a fluctuating condensed film on the surface allowing easy gliding of small water drops as well as generating an internal noise. This scenario deserves careful analysis with adequate theory and experiments. Recent studies on heterogeneous wettability and hierarchical roughness[68], evaporation induced directed motion[69] of a drop on a morphological gradient surface, coalescence induced propulsion of drops[70], and motion induced by chemical reaction[8] are some examples of such systems with which to discover parallels to transport processes found in nature, including an active transport. A drop moving uphill[7] by a chemical force is only a prelude to such vast possibilities.

Generation of motion of liquid droplets in contact with a surface is not only important in various technological settings, it also has a significant philosophical underpinning in that it forms an interesting paradigm for non-equilibrium thermodynamics. For example, the evolution of noise resulting from pinning depinning transitions could be used to understand the physics of critical dynamics[43], in which the development of fluctuation announces whether a system would speed up or slow down critically, long before such a catastrophe actually happens. Barrier crossing experiments with a drop could also serve as a model system with which to study the physics of Kramers' transition involving soft deformable particles. The scope of these studies can be enhanced further by bringing in the role of elasticity either in the drop (e.g. hydrogel) and/or the substrate itself (see SI).

## ACKNOWLEDGEMENTS.

Prof. G. M. Whitesides suggested the original problem of drop motion, with whom the senior author (MKC) had many enjoyable collaborations. We also benefited considerably from many advises received from (late) Prof. P. G. de Gennes. We learned some important issues of critical dynamics from interesting exchanges with Prof. Y. Pomeau.

**Supporting Information:** A list of nomenclatures, supplementary materials and twelve movies. This material is available free of charge via the Internet at http://pubs.acs.org.